\documentclass[twocolumn,epsf,prb,amsmath,amssymb,showpacs]{revtex4-1}
\usepackage{graphicx}
\usepackage{color}
\begin{document}
\title{Carbon clusters: from ring structures to nano-graphene}

\author{D. P. Kosimov$^1$, A. A. Dzhurakhalov.$^{1,2}$, and F. M. Peeters$^1$}
\address{$^1$Department of Physics, University of Antwerp,
Groenenborgerlaan 171, B-2020 Antwerpen, Belgium\\
$^2$Theoretical Department, Arifov Institute of Electronics, Durmon yuli 33, 100125 Tashkent, Uzbekistan}

\begin{abstract}
The lowest energy configurations of $C_{n} (n\leq55)$ clusters are obtained using the energy minimization technique with the conjugate gradient (CG) method where a modified Brenner potential is invoked for the carbon and hydrocarbon interaction. We found that the ground state configuration consists of a single ring for small number of $C$ atoms and multi-ring structures are found with increasing $n$, which can be in planar, bowl-like or cap-like form. Contrary to previous predictions, the binding energy $E_b$ does not show even-odd oscillations and only small jumps are founded in the $E_b(n)$ curve as a consequence of specific type of edges or equivalently the number of secondary atoms. We found that hydrogenation of the edge atoms may change the ground state configuration of the nanocluster. In both cases we determined the magic clusters. Special attention is paid to trigonal and hexagonal shaped carbon clusters and to clusters having a graphene-like configuration. Trigonal clusters are never the ground state, while hexagonal shaped clusters are only the ground state when they have zigzag edges.

\end{abstract}

 \maketitle

\section{Introduction}

The fabrication of single layers of graphite - known as graphene\cite{novo} has resulted in an intense interest in the structure and properties of stable graphene-like structures. Questions concerning the formation mechanisms and the possible stable structures, especially the most stable ones, has focused attention on small carbon clusters both by experimentalists and theorists. Even with modern probe techniques\cite{bill} it is often difficult to obtain unambiguous structural information from experiment\cite{guo,sheli,heat} for small size clusters. On the theoretical side, the study of carbon clusters can be divided into two main categories: calculations using $ab initio$ techniques\cite{fahy,ragh,galli,fura} and calculations based on empirical interatomic potentials\cite{xu,hobd,zhang,toma}. Accurate $ab-initio$ calculations performed by Raghavachari and Binkley (Ref. 7) obtained the structure and energy of small carbon clusters $C_{n} (n=2-10)$ and predicted an odd-even alternation in the nature of the cluster geometries with the odd-numbered clusters having a linear structure and the even-numbered clusters preferring an irregular cyclic structure.

There are two kinds of empirical interatomic carbon potential functions that are widely used. One is the empirical potential proposed by Brenner, which was based on the Abell-Tersoff bond-order expression and fitted on diamond, graphiti crystals and small hydrocarbons1\cite{bren}. The validity and reliability of the Brenner potential to the study of fullerene, nanotubes and carbon cluster formation have been demonstrated in a range of applications\cite{robe,schw,yama,maru,hobd,petu,wang,mylv}. With such potential function C. Zhang et al. (Ref. 13) used a genetic algorithm (GA) together with a simulated annealing (SA) method to study the geometry of $C_{n} (n=2-30)$ clusters. D. H. Robertson et al.\cite{robe} used molecular dynamics (MD) to obtain three-ring structures up to fullerenes by modifying one of the parameters in the Brenner potential function. Y. Yamaguchi and S. Maruyama\cite{yama}, and S. Maruyama and Y. Yamaguchi\cite{maru} used molecular dynamics (MD) to study the formation of fullerenes out of 500 carbon atoms that were generated randomly in a cubic box by omitting one of the parameters in such a Brenner-type potential function. Y. Wang et al.\cite{wang} used a time-going backward quasi-dynamics method with the Brenner potential in order to optimize cluster structures. K. Mylvaganam et al.\cite{mylv} applied MD to analyze carbon nanotubes and their mechanical properties. W. Cai et al.\cite{cai} used a global optimization algorithm to study structural properties of $C_{n} (n=2-71)$ clusters.

Another kind of potential function is the semi-empirical tight- binding (TB) potential which was based on the extended H\"{u}ckel molecular orbital approximation. Such kind of potentials were used to obtain equilibrium geometries for carbon clusters of arbitrary size\cite{wang2,xu2}.

Recently, magnetic order in different shaped graphene clusters, that arise from size effects or from topological frustration, has been proposed as building elements for logic gates in novel ultrafast high-density spintronic devices that can work at room temperature\cite{wang3}. In these calculations the shape of the clusters and ordening of the carbon atoms in the clusters were assumed to be the same as in the ideal graphene sheet without relaxation of the C-atom positions and without questioning if such clusters can be stabilized or not.

Here, using energy minimization with the conjugate gradient (CG) method, we obtain the stable states of different size clusters and obtain the lowest energy configuration of the carbon atoms for quasi-2D arrangements. We investigated the equilibrium configuration of different size clusters and examined when graphene lattice-type structures can be formed. Both even- and odd-numbered neutral carbon clusters $C_{n}$ were systematically studied in our previous work\cite{kosim} for$(n=2-10)$. All possible stable configurations were found, and several new isomers were predicted. It should be pointed out that the dependence of the binding energy for linear and cyclic clusters versus the cluster size $n$ $(n=2-10)$ was found to be in good agreement with several previous calculations, in particular with $ab-initio$ calculations as well as with experimental data for $C_{n} (n=2-5)$\cite{kosim,ragh2}. In the present paper we extend our previous work and calculated the ground state energies for $C_{n} (n=11-55)$
carbon clusters. A detailed study will be presented of the possible geometries of neutral $C_{n} (n=11-55)$ carbon clusters and we determine the ground state, using the energy minimization technique with the Brenner potential function using the CG method described in our previous work (Ref. 26). This method is used to obtain equilibrium geometries, bond length, and binding energies for carbon clusters of arbitrary size. The number of possible stable isomers increases very rapidly with increasing cluster size, and it is not possible to study all those clusters for $n$ beyond 30 atoms. For this reason in this work we limit ourselves to the lowest energy configurations and pay special attention to hexagonal and trigonal shaped graphene flakes. The effect of passivation of the edge atoms by $H$ on the energy and stability of the clusters is studied in detail.

This paper is organized as follows. In Sec. II, a description is given of the inter-atomic potential used in our calculation. The ground state geometries and energetics of isomers of small carbon clusters $C_{n} (n=11-55)$ predicted in this study are compared with available theoretical and experimental data of others in Sec. III. The effect of H-passivation of the edge atoms on the energy and the configuration is investigated in Sec. IV. The stability of carbon clusters are discussed in Sec. V. The energetics of trigonal and hexagonal shaped clusters are studied in Sec. VI. In Sec. VII we summarize our main results.

\section{simulation method}

The computational technique and the Brenner second- generation reactive empirical bond order (REBO) potential\cite{bren2} function between carbon atoms are similar as those used in our previous work (Ref. 26). Here we will concentrate on the main differences with the approach of Ref. 26. The values for all the parameters used in our calculation for the Brenner potential can be found in Ref. 28 and are therefore not listed here. The Brenner potential (REBO) terms in this work were taken as follows:

\begin{equation}
{E_b}=\sum_{i} \sum_{j(>i)} [V^R(r_{ij})-b_{ij}V^A(r_{ij})]
\end{equation}

Here ${E_b}$ is the average binding energy in electronic volt (eV), $V^R(r_{ij})$ and $V^A(r_{ij})$ are a repulsive and an attractive term, respectively, where ${r_{ij}}$ is the distance between atoms $i$ and $j$ given by

\begin{equation}
V^R(r)=f^C(r)(1+Q/r)Ae^{-\alpha r} 
\end{equation}

\begin{equation}
V^A(r)=f^C(r)\sum_{n=1,3} B_n e^{-\beta_{n} r} 
\end{equation}
where the cut-off function $f^C(r)$ is taken from the switching cutoff scheme 

\begin{equation}
\label{hj}
\displaystyle{f_{ij}^{C}(r)=\left\{\begin{array}{l}
                         1  ~~~~~~~~~~~~~~~~~~~~~~~~~~~~~~~~~~~ r<D_{ij}^{min}\\
                        \big[1+cos\big(\frac{(r-D_{ij}^{min})}{(D_{ij}^{max}-D_{ij}^{min})}\big)\big] /2~~D_{ij}^{min}<r<D_{ij}^{max}\\
                         0 ~~~~~~~~~~~~~~~~~~~~~~~~~~~~~~~~~~~~r>D_{ij}^{max}
               \end{array}
\right.
}
\end{equation}
$A, Q, \alpha, B_n, \beta_n, (1\leq n \leq 3)$ are parameters for the carbon-carbon pair terms. Here $n$ is the type of bonds (i.e. single, double and triple bonds).

The empirical bond order function used in this work is written as a sum of terms:

\begin{equation}
b_{ij}=\frac{1}{2}[b_{ij}^{\sigma-\pi}+b_{ji}^{\sigma-\pi}]+b_{ij}^{\pi}
\end{equation}
where the functions $b_{ij}^{\sigma-\pi}$ and $b_{ji}^{\sigma-\pi}$ depend on the local position and bond angles determined from their arrangement around each atom ($i$ and $j$, respectively) and is governed by the hybridization of the orbitals around the atom. Since in the present work we restricted ourselves to quasi-planar structures, the conjugate � compensation term $F_{ij}(N_i, N_j, N_{ij}^{conj})$ was omitted following Ref. 13, which is now given by

\begin{equation}
b_{ij}^{\sigma-\pi}=\big[1+\sum_{k(\ne i,j)} f_{ik}^C(r_{ik})G(cos(\theta_{ijk}))\big]^{-1/2}
\end{equation}
Here the angular function $G(cos(\theta_{ijk}))$ modulates the contribution of each nearest neighbour and is determined by the cosine of the angle of the bonds between the atoms $i-j-k$.

Some angles between two carbon atoms are less than $109.47^{\circ}$, as for instance in the case of pentagonal rings (Ref. 26). Therefore the revised angular function $g_C$ for $109.47^{\circ}$ and $0^{\circ}$ (6-7) was used

\begin{equation}
g_C=G_C(cos(\theta))+Q(N_i^C)[\gamma_C(cos(\theta))-G_C(cos(\theta))]
\end{equation}
with

\begin{equation}
\label{hj}
Q_i(N_i^C)=\left\{\begin{array}{l}
                         1~~~~~~~~~~~~~~~~~~~~~~~~~~~~~~~~~~~~N_i^C<3.2\\
                          \big[1+cos\big(2\pi(N_i^C-3.2)) /2\big]~~3.2<N_i^C<3.7\\
                         0~~~~~~~~~~~~~~~~~~~~~~~~~~~~~~~~~~~~ N_i^C>3.7
               \end{array}
\right.
\end{equation}
where $N_i^C$ is the number of neighbours of carbon atom $i$. $\gamma_C(cos(\theta))$ is a second order spline which was determined for angles less than $109.47^\circ$ which is coupled to $G(cos(\theta))$ through the local coordination, and it retains the value of first and second derivatives at $\theta=109.47^{\circ}$ of the original function $G(cos(\theta))$. The function $b_{ij}^\pi$ in equation (5) is well described in Ref. 28.

We calculated all possible planar structures which are close to the ground state, and in particular we will concentrate on graphene-like structures. Moreover we studied different graphene structures with various shape and topology in order to learn which of them has the lowest energy, i.e. is the ground state. Furthermore, H-passivation of the edge atoms will also be considered.

\section{CONFIGURATIONS AND ENERGIES OF CARBON CLUSTERS}

The optimized geometries are shown in Fig. 1. Table I lists the energy of the clusters per atom. Their shapes and the corresponding bond lengths between the atoms. Some data from the literature are given in parentheses (Ref. 13) and in brackets (Ref. 22) for the sake of comparison.

As can be seen from Fig. 1 the most stable planar clusters are �close-packed� structures that have mainly zigzag edges. This conclusion is in agreement with previous works.\cite{zhang,wang} We found that for $n<19$ the most stable structure is a mono-ring which contrasts with previous results of Ref. 13 where such a single ring configuration was found to be the ground state only up to $C_{12}$. R.O. Jones\cite{jone} used the density functional technique and found mono-ring structures up to $C_{16}$. M. Sawtarie et. al.\cite{sawt} investigated the relative stabilities of ring, multi-ring and cage structures of $C_{20}$ clusters. Raghavachari et. al.\cite{ragh} considered a mono-ring structure (with $D_{2h}$ symmetry) for $C_{24}$. C.H. Xu et. al obtained mono-ring structures up to $C_{19}$.\cite{xu}	D. Tom\'{a}nek\cite{toma} obtained, by combining an adaptive simulated annealing method and a simple tight-binding-type Hamiltonian for the total energy, mono-ring structures up to $n$=22. These different maximum $n$ values for which a mono-ring is the ground state is a consequence of the different theoretical formalisms that have been used. It also implies that the accuracy with which the sp-hybridization is described is crucial in determining the stability of these carbon nanorings for $n>11$.

In Fig. 2(a) we plot the binding energy per atom as function of the number of atoms in the cluster and compare it with other calculations and with available experimental data. Notice that the experimental and $ab-initio$ results are close to our results which exhibit somewhat less scatter. For $n>5$ our results start to deviate from the GA/SA data which are lower in energy. The latter is a consequence of the different interatomic potential (i. e. earlier generation of Brenner potential) that was used (the same remark holds for the results of Ref. 22). We notice pronounced oscillations in the GA/SA results for $n>15$ which are related to the shape of the clusters and the position of the polygons inside certain clusters. These oscillations are absent in our results and in those of Wang et al.\cite{wang} This casts doubts on the accuracy of the energies found from the genetic algorithm of Ref. 13. Fig. 2(b) shows our results (full squares) for the binding energy over the full range of investigated cluster sizes. We can divide Fig. 2(b) in five zones. In the first zone we found single ring clusters $C_{11- 18}$ which are structures having $D_{nh}$ $(n = 11 - 18)$ symmetry without variations in the bond lengths. The bond length decreases with increasing n and lies in the range 1.348-1.338 \r{A}, which is typical for a double-bond, and which compares with the bond lengths 1.334-1.338 \r{A} for $C_{11}$ and 1.333-1.336 \r{A} for $C_{12}$ obtained by the GA/SA technique. In Ref. 24 these mono-ring structures with the same symmetry, were predicted up to $n < 17$ with the binding energy of $C_{17}$ equal to -6.1668 eV/atom which compares with our result -6.0093 eV/atom.

The reason for the stability of such mono-ring configurations for large $n$ was given by M. Sawtarie et al.\cite{sawt} The larger ring diameter of single ring structures results in a large angle between adjacent $C-C$ bonds which minimizes the overlap of $\pi$ electrons in the plane of the carbon atoms, that would otherwise have resulted in a symmetry lowering through in-plane distortions.

The second zone starts with a drop in the derivative of the binding energy where the ground state structure transforms from a single ring to a planar graphene-like structure. $C_{19}$ is a structure with five hexagons, which forms zigzag edges with the average binding energy -6.2220 eV/atom. There are small variations in the bond lengths, i. e. 1.371-1.451 \r{A}, which compares with a bond length of 1.42 \r{A} for bulk graphene. Thus, the transformation from a single ring structure to a five hexagon graphene structure causes a drop in the binding energy.

\begin{figure} \vspace*{0.9cm}\centering {\resizebox*{!}{10.2cm}
{\includegraphics{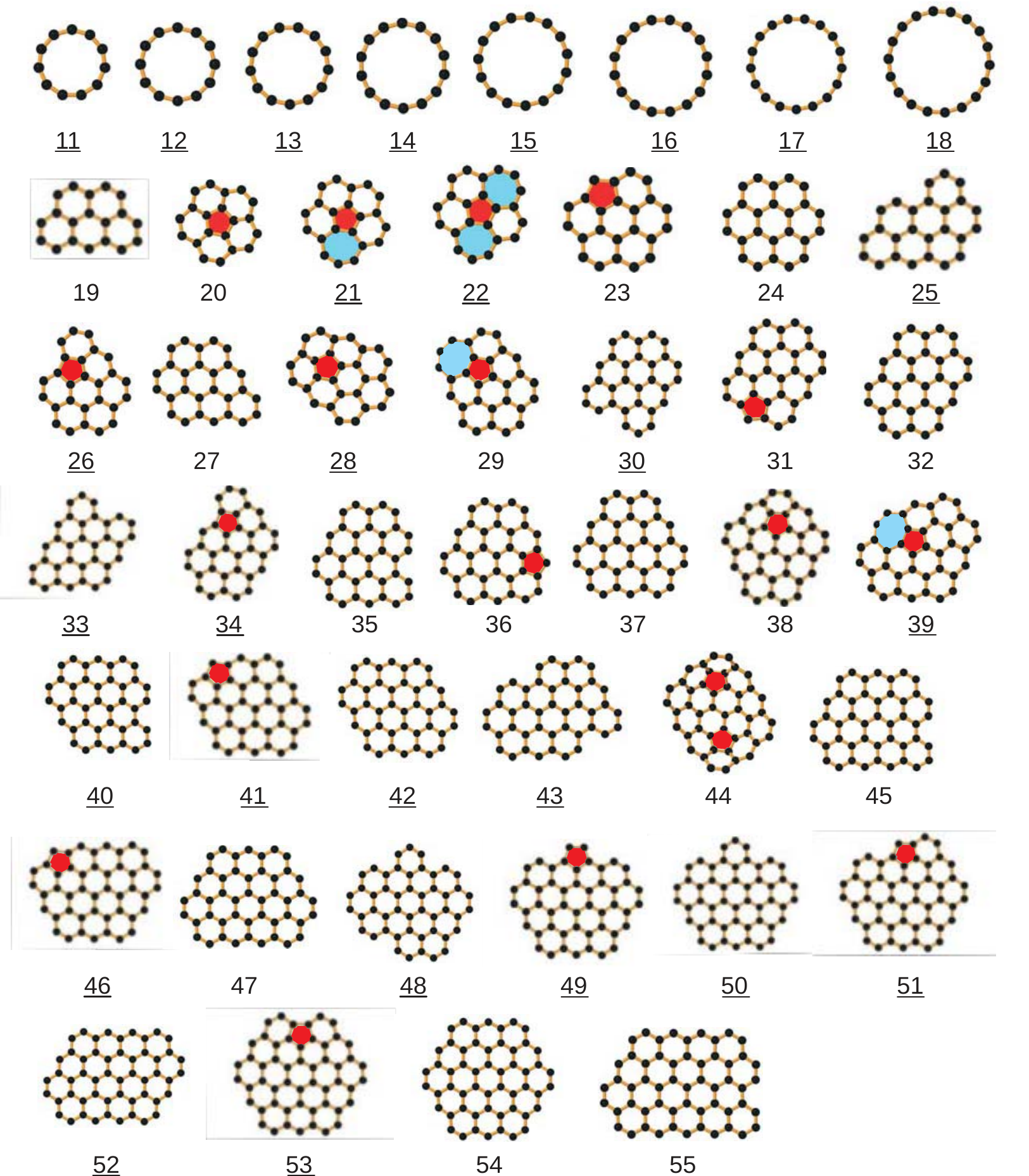}}} \caption{
The ground state configurations for $C_n (n=1-55)$ clusters. The underlined numbers are for clusters whose configuration changes when H-passivated. Pentagons (heptagons) are coloured red (blue).
}\label{fig1}
\end{figure}

The ground state the $C_{20}$ cluster consists of a bowl-like structure, which agrees with the result of Ref. 13. It has one central pentagon and five hexagons surrounding it, with bending $|z|_{max}$=0.965 \r{A} (see Fig. 3). The bond lengths have a smaller variation as compared to those of $C_{19}$ (Table I), with a maximal difference of 0.061 \r{A}. We found that $C_{11}$ and $C_{20}$ carbon clusters can form metastable structures with rhombic polygons. In the case of $C_{21}$ the most stable structure is the planar multi-ring isomer which is constructed from one central pentagon, surrounded by four hexagons and one heptagon. $C_{22}$ is similar to $C_{21}$ where one hexagon is replaced by a heptagon. Notice that these heptagons in the structure prevent the bending of small clusters with a pentagon in the centre. Such clusters were obtained by GA/SA for $C_{20}$ (Ref. 13), and by the TQM method for the $C_{21}$ and $C_{22}$ configurations. The $C_{23}$ and $C_{24}$ are small hexagonal shaped structures. $C_{24}$ consists of purely seven hexagons while $C_{23}$ has a defect, i.e. a pentagon at the edge. The extra atom that sits on the top of the armchair edge is responsible for an ac(56) edge reconstruction.\cite{kosk} Thus $C_{23}$ has one hexagon less than $C_{24}$. The bond length differences in the structure are 0.093 \r{A} for $C_{23}$ and 0.058 \r{A} for $C_{24}$. Since, the reconstructed edge of $C_{23}$ is symmetric, the bond lengths between armrest atoms in the edge is 1.417 \r{A}, and bond length of the C-atom with dangling bond is 1.469 \r{A}. The second zone finishes at the $C_{24}$ cluster. $C_{24}$ is a planar graphene structure and it is the evolutionary form of $C_{23}$ with seven hexagonals and a maximum bond length difference of 0.058 \r{A}. The $C_{24}$ cluster is hexagon shaped with zigzag edges. One can divide the structure into two types of six C-atom hexagons: inner and outer one. The inner polygon�s bond lengths are 1.443 \r{A}, the outer one�s are in the range 1.399-1.391 \r{A}, the most outer bonds� lengths, which involve both sub-lattice atoms, are shorter in comparison with the others.

\begin{figure} \vspace*{0.9cm}\centering {\resizebox*{!}{14.2cm}
{\includegraphics{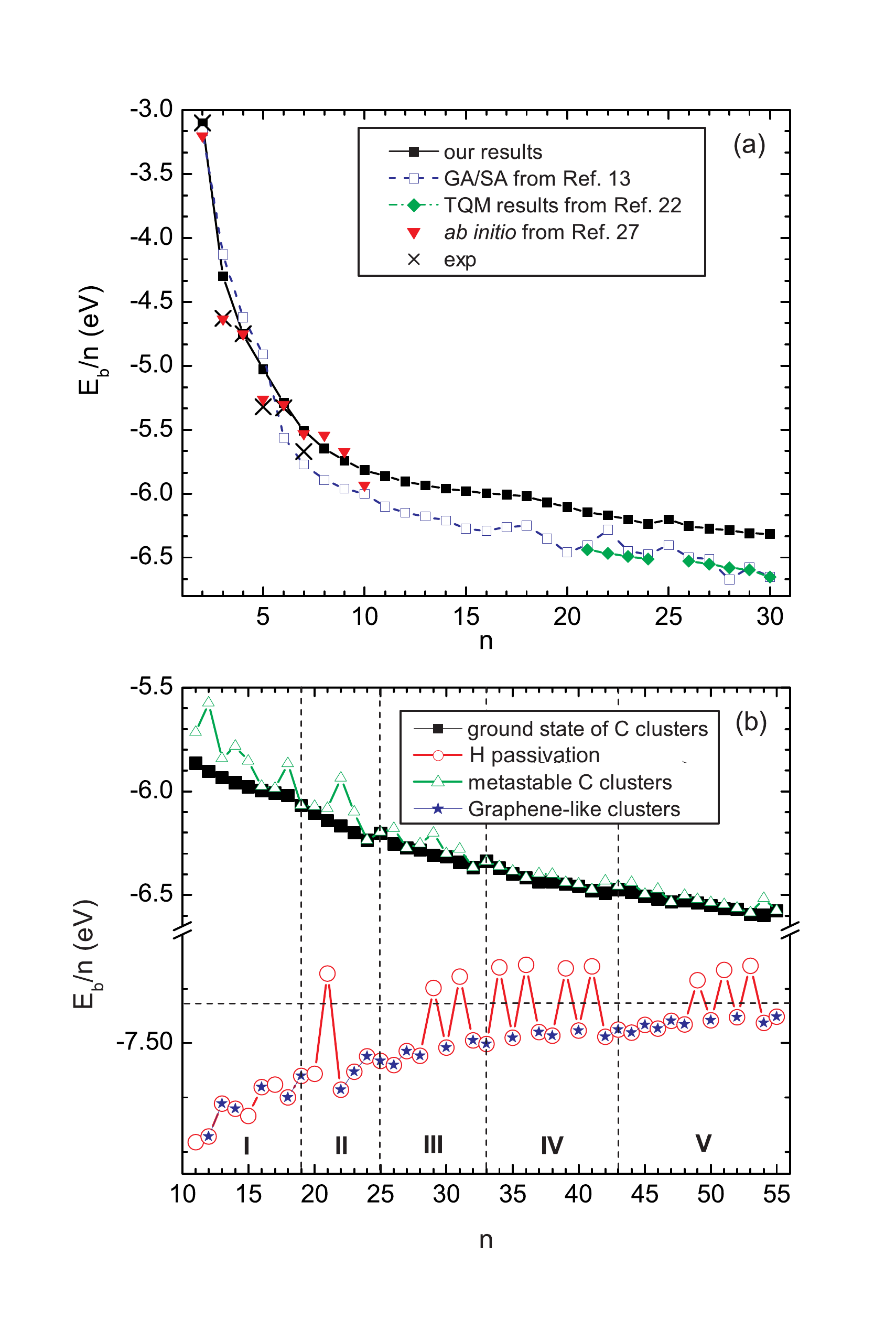}}} \caption{
The binding energy $E_b$ per atom versus the number $n$ of C-atoms in the clusters. (a) Comparison between our results and those from GA/SA, experimental data, $ab-initio$, and TQM calculations. (b) The red circles are the binding energy of carbon atoms for ground state structures with H-passivation, and the open triangles are the corresponding (metastable) structures without H-passivation. The ground state energy of the pure C-clusters is given by the solid square symbols which can be divided in groups as shown by the vertical dashed lines. The horizontal dashed line separates the H-passivated clusters into two groups.
}\label{fig2}
\end{figure}

The bonds that connect the inner ring with the outer one have a length of 1.499 \r{A} and are the longest bonds in this graphene structure. The average binding energy of the edge atoms in $C_{24}$ is -5.8677 eV/atom.

The binding energy increases at the $C_{25}$ graphene structure. Its ground state consists of seven hexagons as $C_{24}$, but with a different arrangement of polygons. Thus $C_{24}$ is a close packed structure, which purely consists of hexagons, while the hexagons in $C_{25}$ are far removed from each other. $C_{25}$ has two type of edges: four zigzag and one armchair edge. It is the first structure which has two different type of edges (the $C_{19-24}$ clusters have only zigzag edges). The bond length differences in the cluster is 0.081 \r{A}.

\begin{figure} \vspace*{0.9cm}\centering {\resizebox*{!}{7.7cm}
{\includegraphics{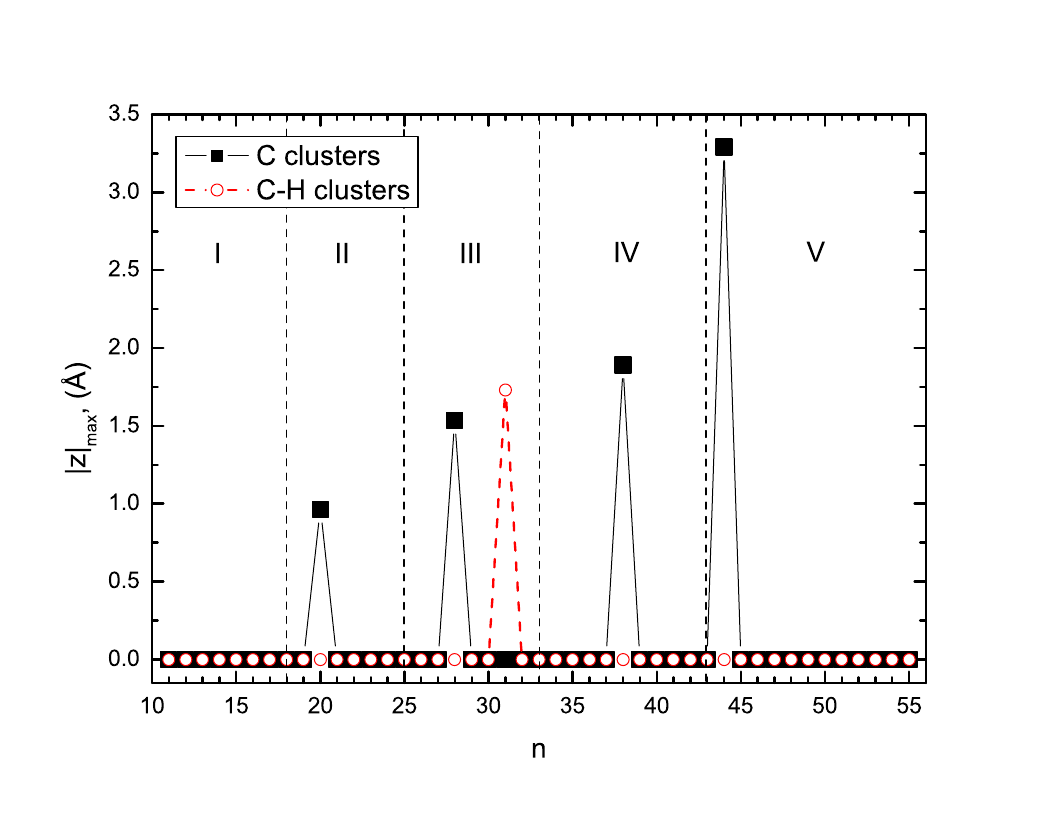}}} \caption{
Dependence of the cluster�s bending $|Z|_{max}$ (\r{A}) as function of the cluster size $n$.}\label{fig3}
\end{figure}

The planar $C_{26}$ and $C_{27}$ clusters are evolutionary forms of $C_{23}$ and $C_{24}$ structures. $C_{26}$ has a defect due to the removal of one atom from the zigzag edge of $C_{27}$, and thus it has one pentagon. $C_{27}$ consists of eight hexagons and it is a graphene-type structure. The bond length differences for $C_{26}$ are 0.099 \r{A} and 0.08 \r{A} for $C_{27}$. As in the previous structures the longest bond length corresponds to the outer bonds with average bond length 1.446 \r{A}.

$C_{28}$ and $C_{29}$ are the evolutionary forms of the $C_{20}$ and $C_{21}$ clusters, which are obtained by adding six atoms to the edge. We have a cap-like $C_{28}$ and planar $C_{29}$ configurations (see Fig. 3), with the bond length differences of 0.075 \r{A} and 0.078 \r{A}, respectively. The bending, of $C_{28}$ is $|z|_{max}$=1.535 \r{A}. The longest bond lengths appear in the pentagon with average length of 1.446 \r{A} as in $C_{26}$. The pentagon in $C_{29}$, like in $C_{21}$, is surrounded by four hexagons and one heptagon. The outer torsion bonds of the heptagon have the shortest bond length, i.e. 1.370 \r{A}. The next structures with the number $n$ = 30 - 33 are planar, evolutionary forms of each others. Their basic configuration is $C_{24}$. All of them are, graphene-type clusters, except $C_{31}$, which exhibits an ac(56) type of edge reconstruction. Since the reconstructed edge is symmetric as in $C_{23}$, the bond lengths between armrests atoms is 1.418 \r{A} and for the dangling bonds it is 1.471 \r{A}. The bond lengths in $C_{30}$ differ at most by 0.079 \r{A}. There are two types of edges in the structure: armchair and zigzag. The length of the cluster along the armchair edge is 6.057 \r{A} and along the zigzag edge is 5.983 \r{A}. Nevertheless the middle width length of the structure is 6.964 \r{A} that means, the cluster is bended inside by zigzag edges. The bond length differences in $C_{31}$ is 0.101 \r{A}. The length of the cluster is 7.489 \r{A} and the widths are 5.265-7.003-5.734 \r{A}. This cluster is bended to the outside. The average energy per atom for the edge with pentagon is -5.7242 eV/atom, and for the edge with hexagon is - 5.8751 eV/atom. The bond lengths difference in $C_{32}$ is 0.08 \r{A}. The width of the cluster is 5.712 \r{A} and length is 7.397 \r{A}, and it is bended toward the outside like $C_{31}$. All edges are of the zigzag type with average binding energy -5.92 eV/atom.

\begin{figure} \vspace*{0.9cm}\centering {\resizebox*{!}{14.2cm}
{\includegraphics{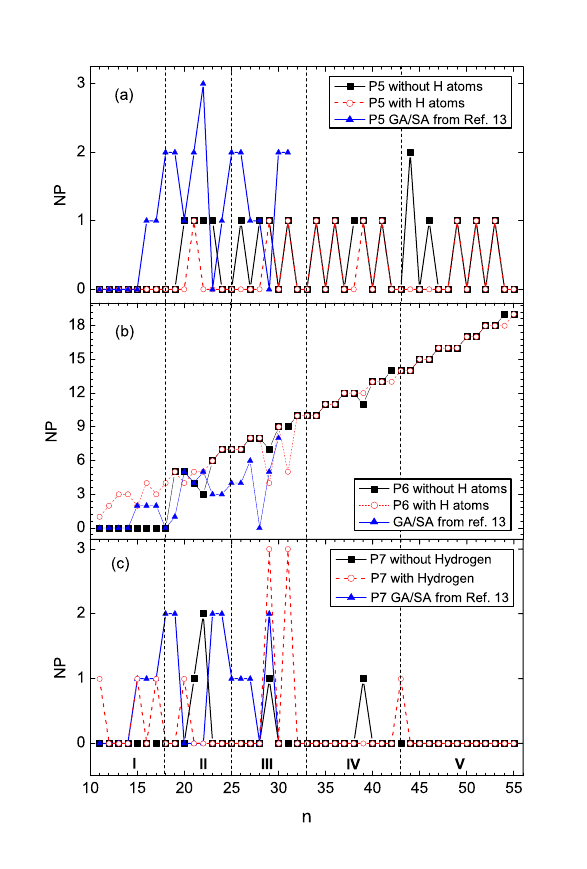}}} \caption{
Comparison between our results and those from GA/SA (Ref. 13) for the number of polygons: � (a) pentagon, (b) hexagon and (c) heptagon in the cluster. Open triangles are those for H- passivated clusters.}\label{fig4}
\end{figure}

\begin{figure} \vspace*{0.9cm}\centering {\resizebox*{!}{19.2cm}
{\includegraphics{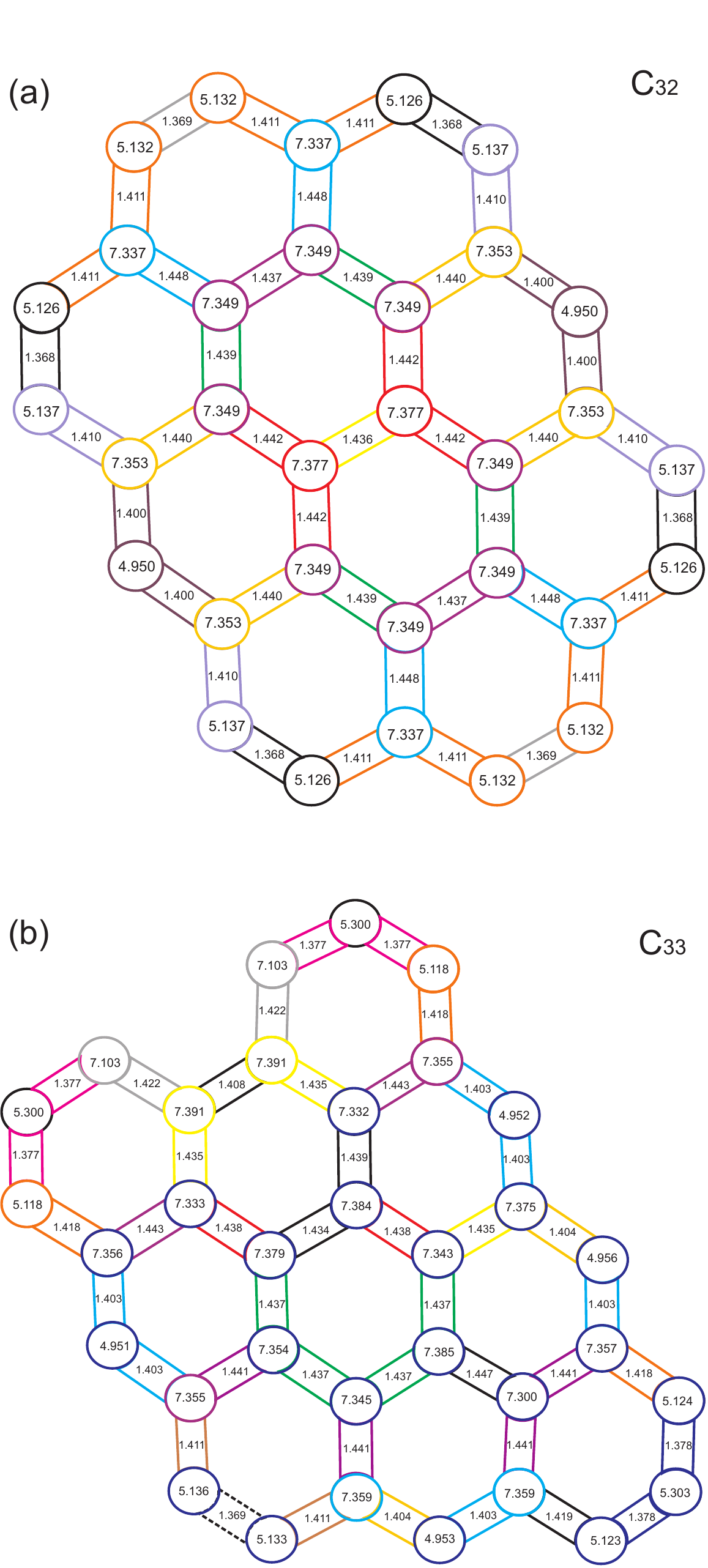}}} \caption{
Spatial configuration of the (a) $C_{32}$ and (b) $C_{33}$ clusters with atomic energies (in eV/atom) and bond length (in \r{A}). The same colors are used for atoms and bonds having practically the same value for the energy and bond length, respectively.}\label{fig6}
\end{figure}

The binding energy exhibits a jump of 0.0311 eV/atom at $n$ = 33 were region IV starts. The bond length differences in $C_{33}$ is 0.078 \r{A}. $C_{33}$ has two types of edges: one armchair and three zigzag edges. Two of the zigzag edges have the same size and they are shorter than the other one. The length of the cluster along the armchair edge is 5.712 \r{A} and along the zigzag edges are 7.417-9.956 \r{A}. The cluster has a trapezoidal shape.

To understand better the jump in the binding energy per atom when we go from $C_{32}$ to $C_{33}$ we show in Fig. 5 the detailed spatial configuration of the atoms, together with the absolute value of the energy of each C-atom in the cluster and the distance between the C-atoms. Notice that ternary atoms (i.e. atoms with three carbon niegbours) have a larger energy, i.e. $E\approx7.35$ eV, which depends slightly on the exact positions in the cluster. Those atoms are connected with three covalent bonds of length 1.44 \r{A}. At the edge we can have secondary atoms that have a substantial lower energy, typically about 5.13 eV with shorter bond length, of about 1.40 \r{A}.

The cluster $C_{32}$ has only zigzag edges with in total 22 edge atoms of which 14 are secondary atoms (i.e. atoms with only two bonds). The other cluster $C_{33}$ has both zigzag and armchair edges with in total 24 edge atoms of which 15 are secondary atoms. Thus when we go from $C_{32}$ to $C_{33}$ the number of edge atoms increases with one. It is the latter increase that is responsible for the smaller average binding energy per atom of the cluster.

Here, a small comment is in order about the two types of evolutionary forms of small clusters that we have found. For instance, the ground state configuration of the clusters $C_{11-18}$, $C_{20-22}$, $C_{23,24}$, $C_{26-33}$, $C_{35-37}$, $C_{40-42}$, $C_{46,47}$, and $C_{49,50}$ are obtained by adding only one atom to a certain ground state configuration. The other type of configurations result from an increase of polygons, such as the structures $C_{20}$, $C_{23}$, $C_{26}$, $C_{34}$, $C_{38}$, and $C_{44}$. Thus, we can distinguish them as atomic and polygonal evolutions. One of those evolutions leads to the creation of planar structures, when the pentagon as a defect is inserted at the edge resulting in an edge reconstruction. The other type of evolution leads to the creation of bended structures, when the pentagon is in the centre of the clusters (see Figs. 1 and 3). So $C_{34}$, in addition to $C_{26}$ has additional three hexagons. The structure�s bond length differences of 0.11 \r{A}. This large difference is due to the pentagon in the cluster with bond length 1.444 \r{A}. All edges in $C_{34}$ are of zigzag type.

The next clusters, i.e. $C_{35-37}$, are atomic evolutionary forms and are all planar configurations. Two of them, $C_{35}$ and $C_{37}$ consist of 11 and 12 hexagons, respectively, with bond length differences of 0.081 \r{A} for $C_{35}$ and 0.08 \r{A} for $C_{37}$ as in $C_{32}$. $C_{35}$ has two types of edges: armchair and zigzag. $C_{37}$ has only zigzag edges. The average binding energy $E_b$ of the edge atoms in the cluster is -5.9380 eV/atom. $C_{36}$ in addition to its eleven hexagons has one pentagon with an armchair ac(56) type of edge reconstruction. The difference of bond lengths in this structure is 0.103 \r{A}. This large difference between bonds is due to the pentagon in the cluster (i.e. as in $C_{35}$) which has an average bond length of 1.453 \r{A}. The $C_{38}$ structure is the polygonal evolutionary form of $C_{20}$, with additional seven hexagons. The bond length differences in this structure is of 0.078 \r{A}. All edges are of the zigzag type, and the structure is bended by $|z|_{max}$=1.891 \r{A}.	$C_{39}$ structure is a unique cluster with one heptagon, having a planar structure with bond length differences of 0.087 \r{A} and only zigzag edges. The $C_{40}$ structure is a planar one, and it is a polygonal evolutionary form of $C_{37}$ with one armchair edge. Thus, $C_{40}$ has two types of edges. The bond length differences is 0.081 \r{A}. The two planar configurations $C_{40}$ and $C_{41}$ are atomic evolutionary forms of each other. $C_{41}$ has one pentagon on the edge, and thus has only zigzag edges. The bond length differences are 0.104 \r{A} for $C_{40}$, and 0.079 \r{A} for $C_{41}$. Adding one atom to the pentagon of the reconstructed ac(56) edge of $C_{41}$ we obtain a �close packed� structure as the ground state having only zigzag edges (see Fig.1). The structure�s bond length difference is 0.079 \r{A}. The fourth zone finishes with the $C_{42}$ cluster, where the average binding energy of the edge atoms is -5.9538 eV/atom.

\begin{table*}
\caption{\label{table1}The binding energy per atom as function of the number of atoms in the (C-C) and (C-H) clusters (the numbers between parentheses are from Ref. 13, and those between brackets are from Ref. 24). The cluster shape is labeled �$SR$� for single rings, �$G$� for graphene, �$G_{cp}$� for close packed, �$DR$� for double ring, �$MR$� for multi-ring, �$B$� for bended structures. The fourth column gives the range of bond lengths appearing in the cluster (results between parentheses are those from Ref. 13). The last three columns are the equivalent results for the H-passivated carbon clusters.}
\begin{ruledtabular}
\begin{tabular*}{1.\textwidth}{@{\extracolsep{\fill}}llclccc}
n & ~~~$E_b^{C-C}$ & Shape & $D_{C_C}$ & $E_B^{C_H}$ & Shape & $D_{C_H}$\\ 
~~~~~&(eV/atom)   & (C-C)   & (\r{A})         & (eV/atom) & (C-H)   & (\r{A}) \\ [0.44ex]
\hline
11 & -5.8647 		   & $SR$        & 1.348             	      & -7.5911 & $DR$	& 1.377-1.428 \\
12 & -5.9041 		   & $SR$        & 1.346             	      & -7.5857 & $DR$	& 1.384-1.415 \\
13 & -5.9341 		   & $SR$        & 1.344             	      & -7.5557 & $G_{cp}$ & 1.382-1.428 \\
14 & -5.9595 		   & $SR$        & 1.342             	      & -7.5606 & $G$ 		& 1.383-1.430 \\
15 & -5.9794 		   & $SR$        & 1.341             	      & -7.5671 & $MR$	& 1.373-1.444 \\
16 & -5.9957 		   & $SR$        & 1.340             	      & -7.5406 & $G_{cp}$ & 1.381-1.427 \\
17 & -6.0093 		   & $SR$        & 1.339             	      & -7.5382 & $MR$	& 1.372-1.450 \\
18 & -6.0207 		   & $SR$        & 1.338 	      	      & -7.5498 & $G$ 		& 1.383-1.430 \\
19 & -6.0681                 & $G_{cp}$ & 1.371-1.451    	      & -7.5299 & $G$ 		& 1.381-1.427 \\
20 & -6.1041 (-6.475) & $MR$        & 1.390-1.451 	      & -7.5282 & $MR$	& 1.371-1.449 \\
21 & -6.1424 [-6.44]    & $MR$       & 1.367-1.457  	      & -7.4361 & $MR$	& 1.383-1.443 \\
22 & -6.1674 [-6.465]  & $MR$       & 1.360-1.458 	      & -7.5429 & $G$ 		& 1.380-1.427 \\
23 & -6.1992 [-6.49]   & $MR$        & 1.376-1.469 	      & -7.5265 & $G$ 		& 1.381-1.429 \\
24 & -6.2368 [-6.51]   & $G_{cp}$  & 1.391-1.449 	      & -7.5121 & $G_{cp}$ & 1.381-1.426 \\
25 & -6.2013               & $B$ 	     & 1.370-1.451 	      & -7.5162 & $G$ 		& 1.381-1.428 \\
26 & -6.2533 [-6.525] & $MR$  	     & 1.378-1.477 	      & -7.5200 & $G$ 		& 1.380-1.428 \\
27 & -6.2727 [-6.55]   & $G$ 	     & 1.369-1.449 	      & -7.5075 & $G$ 		& 1.381-1.426 \\
28 & -6.2834 [-6.58]   & $B$ 	     & 1.376-1.451 	      & -7.5116 & $G$ 		& 1.380-1.427 \\
29 & -6.3083 [-6.65]   & $MR$ 	     & 1.370-1.448 	      & -7.4491 & $MR$	& 1.363-1.489 \\
30 & -6.3148               & $G$ 	     & 1.369-1.448 	      & -7.5039 & $G$ 		& 1.381-1.426 \\
31 & -6.3429               & $MR$ 	     & 1.370-1.471 	      & -7.4390 & $B$ 		& 1.368-1.457 \\
32 & -6.3684 		 & $G_{cp}$  & 1.368-1.448 	      & -7.4974 & $G_{cp}$ & 1.381-1.426 \\
33 & -6.3373 		 & $G$ 	     & 1.369-1.447 	      & -7.5010 & $G$ 		& 1.381-1.428 \\
34 & -6.3704 		 & $MR$ 	     & 1.372-1.482 	      & -7.4302 & $MR$	& 1.383-1.462 \\
35 & -6.3963 		 & $G$ 	     & 1.367-1.448 	      & -7.4952 & $G$ 		& 1.381-1.426 \\
36 & -6.4180 		& $MR$ 	     & 1.371-1.474 	      & -7.4280 & $MR$	& 1.383-1.463 \\
37 & -6.4380 		& $G_{cp}$   & 1.368-1.448 	      & -7.4899 & $G_{cp}$ & 1.381-1.426 \\
38 & -6.4365 		& $B$ 	     & 1.374-1.452 	      & -7.4933 & $G$ 		& 1.381-1.426 \\
39 & -6.4473 		& $MR$ 	     & 1.370-1.457 	      & -7.4311 & $MR$	& 1.383-1.470 \\
40 & -6.4575 		& $G$ 	     & 1.367-1.448 	      & -7.4885 & $G$ 		& 1.381-1.426 \\
41 & -6.4748 		& $MR$ 	     & 1.370-1.474 	      & -7.4294 & $MR$	& 1.383-1.464 \\
42 & -6.4913 		& $G_{cp}$   & 1.368-1.447 	      & -7.4943 & $G$ 		& 1.381-1.476 \\
43 & -6.4745 		& $G$ 	     & 1.368-1.447 	      & -7.4875 & $G$ 		& 1.381-1.426 \\
44 & -6.4860 		& $B$ 	     & 1.378-1.454 	      & -7.4905 & $G$ 		& 1.381-1.426 \\
45 & -6.5054 		& $G$ 	     & 1.368-1.447 	      & -7.4833 & $G$ 		& 1.381-1.426 \\
46 & -6.5195 		& $MR$ 	     & 1.368-1.474 	      & -7.4864 & $G$ 		& 1.381-1.426 \\
47 & -6.5336 		& $G_{cp}$   & 1.367-1.448 	      & -7.4797 & $G_{cp}$ & 1.381-1.426 \\
48 & -6.5263 		& $G$ 	     & 1.368-1.447 	      & -7.4828 & $G$ 		& 1.381-1.426 \\
49 & -6.5393 		& $MR$ 	     & 1.368-1.447  	      & -7.4420 & $MR$	& 1.381-1.426 \\
50 & -6.5520 		& $G$ 	     & 1.368-1.447 	      & -7.4792 & $G$ 		& 1.381-1.426 \\
51 & -6.5650 		& $MR$ 	     & 1.369-1.474 	      & -7.4326 & $MR$	& 1.383-1.463 \\
52 & -6.5676 		& $G_{cp}$   & 1.368-1.447 	      & -7.4761 & $G$ 		& 1.381-1.426 \\
53 & -6.5918 		& $MR$	     & 1.372-1.497 	      & -7.4289 & $MR$	& 1.384-1.469 \\
54 & -6.5973 		& $G_{cp}$   & 1.367-1.439 	      & -7.4815 & $G$ 		& 1.381-1.426 \\
55 & -6.5753 		& $G$ 	     & 1.368-1.447 	      & -7.4758 & $G$ 		& 1.381-1.427 \\[1ex]
\end{tabular*}
\end{ruledtabular}
\label{table:nonlin}
\end{table*}%

The $C_{43}$ structure is a planar graphene-type cluster with bond lengths difference of 0.079 \r{A} as in the $C_{42}$ structure. Note that the jump in the binding energy per atom is 0.168 eV/atom which is due to the fact that $C_{43}$	has two types of edges because the average binding energy per atom in the armchair edge is larger than in the zigzag edge. The reason is that armchair termination involves both types of atoms (i. e. both sub-lattice sites). Thus, two types of atoms are close to each other and form pairs. The average bond length in this case is 1.373 \r{A}, which is shorter than the average bond length for zigzag edges (1.41 \r{A}). The $C_{44}$ structure is a polygonal evolutionary form of $C_{20}$. The bond lengths have a difference of 0.076 \r{A} and the cluster is strongly bended with $|z|_{max}$= 3.290 \r{A} (see Fig. 3). All edges are of zigzag type. $C_{45}$ is a planar graphene-type cluster which is a polygonal evolutionary form of $C_{40}$ with the addition of two hexagons. The bond lengths have a difference of 0.079 \r{A}, as in the $C_{42}$ structure, and has two types of edges. The planar $C_{46}$, $C_{47}$, $C_{49}$, and $C_{50}$ configurations are atomic evolutionary forms. $C_{46}$ and $C_{49}$ have an edge defect as pentagon, but with different reconstructions: the edge defect of $C_{46}$ is of ac(56) type while the pentagonal defect in $C_{49}$ was formed by removing one atom from the hexagon. The bond length differences are 0.106 \r{A}, and 0.081 \r{A} respectively for $C_{46}$ and $C_{47}$, and 0.079 \r{A} for both $C_{49}$ and $C_{50}$.

The ground state configurations for $n>50$ are graphene-like clusters. However, $C_{51}$ and $C_{53}$ have a defect on the edge. The defect of the $C_{53}$ cluster is formed due to the lack of one $C$ atom in the hexagon which was positioned in the centre of the zigzag edge. In all cases such kind of ac(56) or ac(67) defects do not lead to a bending of the cluster. The $C_{52}$, $C_{54}$, and $C_{55}$ structures are purely graphene-type clusters. $C_{52}$ and $C_{54}$ consists of zigzag edges, and $C_{54}$ is a hexagon shaped configuration. $C_{55}$ looks like a tiny nanoribbon with armchair and zigzag edges.

Each time the cluster possesses both types of edges the binding energy per atom jumps up (see Fig. 3 (b)). Armchair edges change the cluster structure, and make them extended, rather then �close-packed�. As shown in Fig. 2(b) and Table I the binding energy exhibits a drop when the clusters are �close-packed�, having only zigzag edges. Moreover armchair edges cause an increase of the number of secondary carbon atoms (see Fig. 8). Thus, when we have a larger number of atoms with only two neighbours it results in a smaller binding energy. We found that 42\% of the lowest energy configurations are graphene-type structures.

It is clear from Fig. 4, that the number or the type of polygons in the clusters can not be the cause of oscillations in the binding energy. The oscillations are due to the position of the polygons in the structure, and whether the structure has zigzag or armchair edges.

9 \% of the investigated clusters are bended structures and in all of them there are pentagons in the centre (see Figs. 1 and 3). Those pentagons are surrounded by hexagons and bend them such that their bond length corresponds to a double bond. The smallest bended cluster in the range $n$=11-55 is the $C_{20}$ cluster and the largest one is $C_{44}$ (see Fig. 3). All the clusters after $C_{44}$ are planar configurations. Notice that the buckled structures do not affect the binding energy oscillation.

In Figs. 4 (a, b, c) we plot the number of different polygons, i.e. pentagons (P(5)), hexagons (P(6)), and heptagons (P(7)), constituting the cluster as function of the size of the cluster. Notice that the number of hexagons steadily increases while the number of pentagons and heptagons are mostly one and occasionally two. In the theoretical results of Ref. 13 the pentagons and heptagons play a more important role as compared to our work. The number of pentagons in our results are substantially smaller than in the results of Ref. 13, except for the $C_{44}$ cluster in our work, and except $C_{21}$ in the GA/SA work (both are buckled). However, in our results, starting from $C_{22}$ it is clear that the hexagons dominate which at the end will result in graphene-like clusters (see Fig. 4(b)). We found that clusters which mainly consist of pentagons and heptagons are energetically not favorable, i. e. they are metastable configurations. The single heptagon ring is most stable with average energy per atom -5.5081 eV/atom, which compares with the single pentagon with average binding energy -4.742 eV/atom (see Ref. 26). However in non mono-ring structures the heptagons are positioned at the outer shells while the pentagons are located either in the center of the cluster, or at the outer shell. The reason is that the heptagon has a smaller bond length (1.352 \r{A}) which is close to the double bond length, which compares with pentagons which have 1.437 \r{A} (see Ref. 26). Both, our results and those from the GA/SA work find that hexagon rings play a dominant role in the ground state of clusters.

\section{EFFECT OF H PASSIVATION OF THE EDGE ATOMS}

Often edge atoms are passivated by hydrogen atoms, which will remove the dangling bonds. To our knowledge, the effect of $H$ atoms on the structure of carbon clusters has not been studied yet. In what follows we perform a systematic study of the influence of $H$ passivation of the edge atoms on the stability of carbon clusters.

We attach $H$ atoms on the edge $C$ atoms of all ground state, and metastable configurations which we studied in previous section. From Fig. 2(b) and Table I we see that the binding energy $E_b$ of hydrogen passivated clusters is larger and decreases with increasing $n$, while the binding energy of the pure C-clusters is a decreasing function of $n$. Both curves converge to $E_0=-7.35$ eV/atom in the limit $n\to \infty$. We found that our results for the pure $C$ clusters could be fitted to $E=E_0+\alpha_c / n^{0.5}$ with $\alpha_c$=5.68 eV/atom and for the H-passivated clusters with $\alpha_c$= - 0.77 eV/atom, where we did the fitting over the range $17<n<56$. The square root dependence is a consequence of the fact that the number of edge atoms is proportional to $n^{0.5}$. The binding energy exhibits oscillations as function of n around these fitted curves.

The horizontal dashed line in Fig. 2(b) divides the $H$ passivated clusters into two regions. Below this line we have graphene type clusters (see Fig. 6) with the exception of $C_{11}$, $C_{15}$, $C_{17}$ and $C_{20}$ which all have one heptagon. Above this line those clusters have a defect at the edge, i. e. one pentagon. Notice that the ground state configuration of the small clusters consists no longer of a mono-ring but are mostly graphene-like clusters. We found double, triple, and quarted clusters of hexagons. In the case of pure carbon cluster calculations these structures were metastable configurations (see Fig. 2). $H$ passivation turned them into ground state configurations.

The ground state configurations with hydrogen passivated edge atoms are depicted in Fig. 6. The shape together with the shortest and the longest bond length as given in Table I. Notice that the $H$ passivated clusters have a smaller range of bond lengths as compared to the clusters without $H$ passivation. Similar as on previous section, we have two kinds of evolutionary forms of structures with increasing size.

The non-graphene-like structures can be dividing into two types by the position of the pentagon defect. In the first type the defect is formed by the lack of a $C$ atom at the edge hexagon. That kind of defect appears in all small $C$ clusters, resulting in a lower binding energy (see Fig. 2). The other type of clusters, have a pentagon defect formed by the diffusion of carbon atoms from distant armrests to �seat� positions, thus leading to an ac(56) reconstruction of an armchair edge (Ref. 40).

The largest cluster we studied, i. e. $n$=55, has the same configuration with and without $H$ passivation and has a graphene structure. The difference between the longest and the shortest bond lengths is 0.079 \r{A} for the pure C-cluster and 0.046 \r{A} for the $H$ passivated cluster. The clusters with $H$ passivation are slightly compressed as compared to the pure carbon clusters.

\begin{figure} \vspace*{0.9cm}\centering {\resizebox*{!}{14.5cm}
{\includegraphics{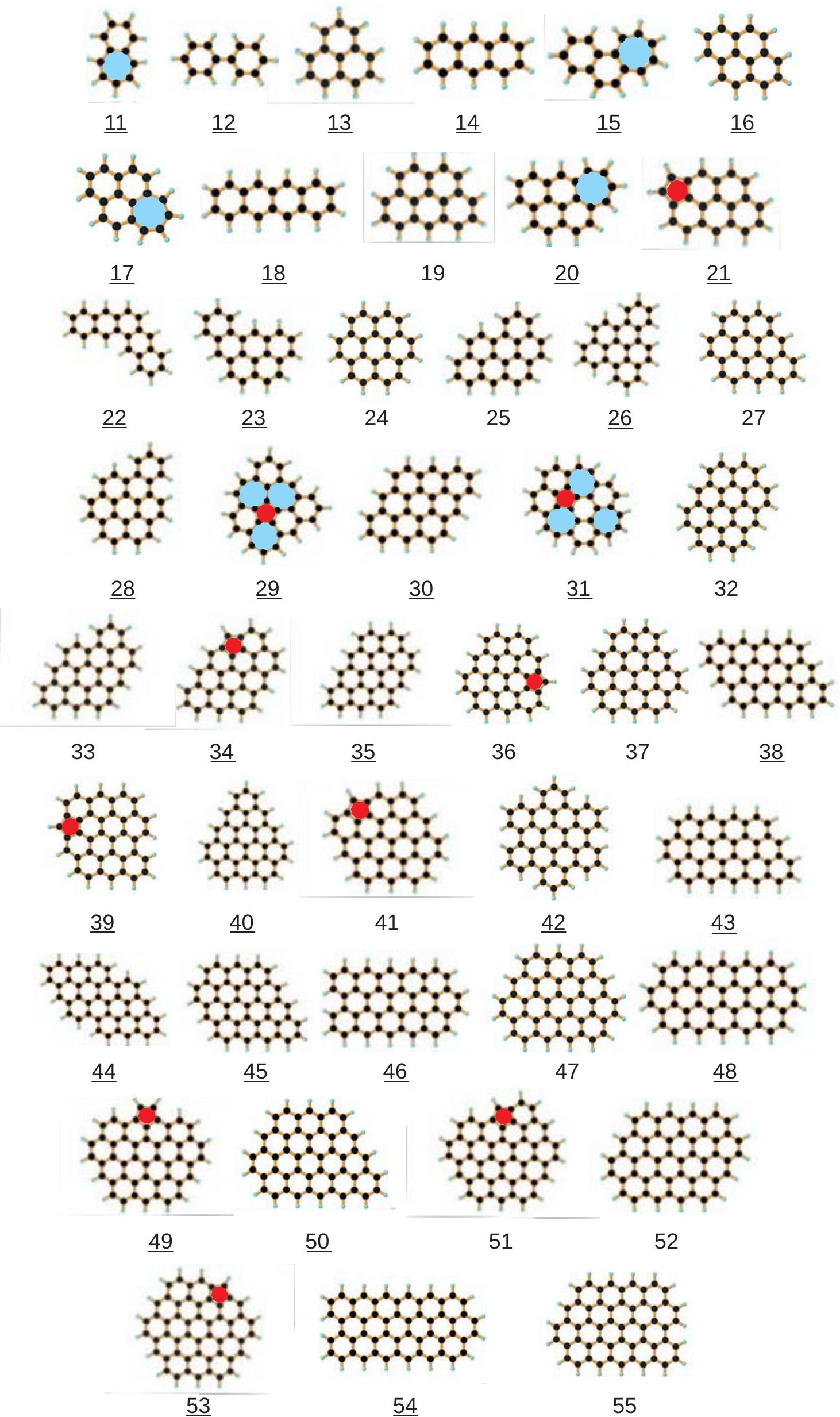}}} \caption{
The ground state configurations for $H$ passivated $C_n (n=11-55)$ clusters. The underlined numbers are clusters whose configuration is different from the one of the pure $C$ clusters that are shown in Fig. 1. Pentagons (heptagons) are colored red (blue).}\label{fig8}
\end{figure}

69\% of the hydrogen passivated structures are graphene-type clusters, and only one ground state configuration $C_{31}$ is a bended cluster.

\section{MAGIC CARBON CLUSTERS}

In order to find the magic clusters we calculate the second finite difference of the binding energy per atom, $\Delta2E(n)=E(n-1)+E(n+1)-2E(n)$, which are shown in Figs. 7(a,b). The positive peaks indicate particularly stable clusters compared to their neighbouring structures. For both C-C and C-H clusters there is no relation between the stability of the cluster and their even- or odd numbered constructions, as was found in the case of 3D structures (Ref. 14). In Fig. 8(a) we can see that the small size single ring structures are most stable, and the biggest single ring $C_{18}$ structure is the least stable one. An analogous trend for single ring structures was found in Ref. 24. The graphene-type hexagonal shaped $C_{24}$ cluster is clearly a magic cluster. Note that $C_{24}$, $C_{32}$, $C_{42}$, and $C_{54}$ configurations are polygonal evolutionary forms of the two hexagonal shaped $C_{24}$ and $C_{54}$ clusters. The highly symmetric clusters are the most stable one.

\begin{figure} \vspace*{0.9cm}\centering {\resizebox*{!}{11.0cm}
{\includegraphics{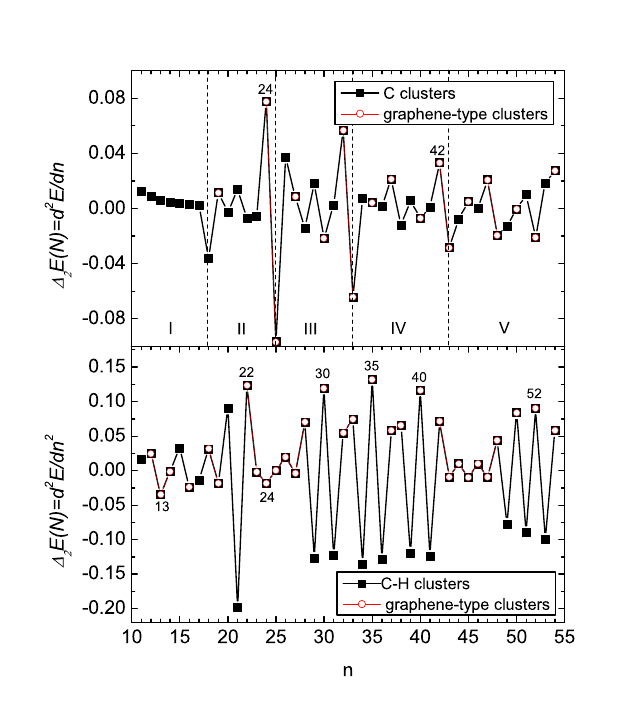}}} \caption{
The second finite difference of the lowest energy $\Delta2E(n)$ per atom vs. cluster size $n$. Peaks correspond to the most stable structures compared to adjacent sizes. (a) is for pure C-clusters and (b) is for C-clusters with H-passivated edges.}\label{fig9}
\end{figure}

Although $C_{25}$, $C_{33}$, $C_{43}$, and $C_{48}$, are graphene-type clusters they have armchair edges, and are the least stable structures. Despite the fact that $C_{52}$ has only zigzag edges, it has one extra edge atoms in contrast to its neighbours. Thus, this additional atom increases the number of secondary carbon atoms and decreases its stability.

In the case of H-terminated clusters we see from Fig. 7(b) that $C_{22}$, $C_{30}$, $C_{35}$, and $C_{52}$ are very stable clusters, i. e. magic number clusters, and are polygonal evolutionary forms of the $C_{22}$ and $C_{52}$ configurations. We can see that the least stable clusters are those having a pentagon.

In order to study the effect of the type of edges on the stability of the clusters we compare the number of zigzag edges in a ground state structure with the number of armchair edges. We found that zigzag edges occur more frequently in both type of ground state clusters (see Figs. 8(a-b)). Some structures exhibit solely zigzag edges, while no ground state clusters where found with only armchair edges (see Fig. 8).

\begin{figure} \vspace*{0.9cm}\centering {\resizebox*{!}{11.0cm}
{\includegraphics{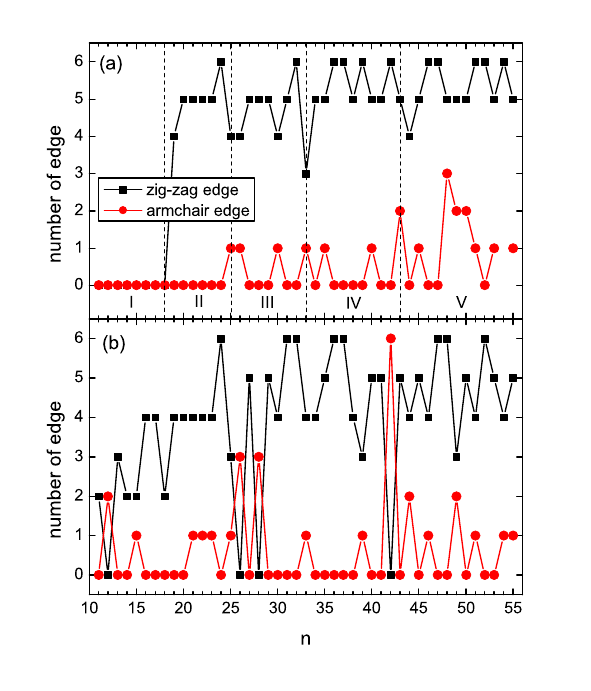}}} \caption{
The number of zigzag and armchair edges in the ground state as function of the size n of the cluster. (a) is for pure C-clusters and (b) is for C-clusters with H-passivated edges.}\label{fig10}
\end{figure}

\begin{figure} \vspace*{0.9cm}\centering {\resizebox*{!}{7.5cm}
{\includegraphics{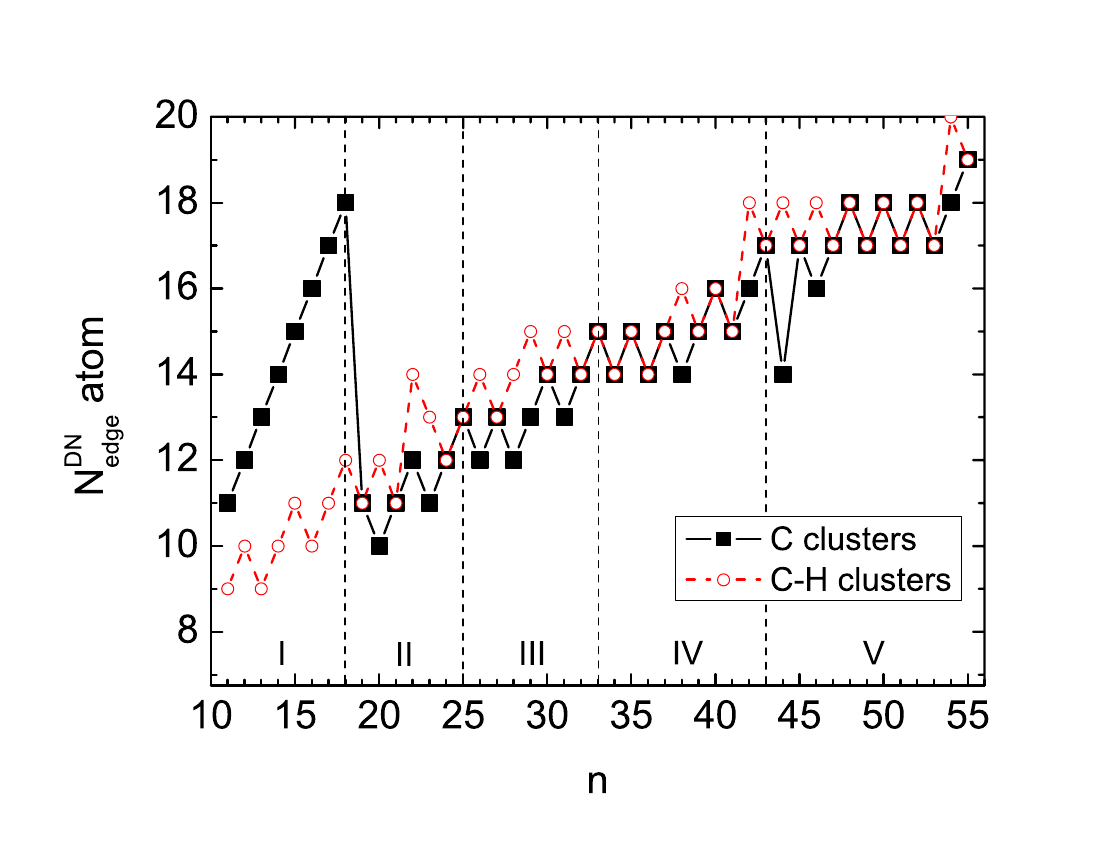}}} \caption{
The number of secondary carbon atoms in the ground state as function of the cluster size $n$.}\label{fig11}
\end{figure}

We can see that each cluster has on the average 5 zigzag edges and for $n<43$ they have 1 or 0 armchair edges for the pure $C$ clusters and a maximum of 3 armchair edges for the $H$ passivated clusters. The number of secondary carbon atoms for both type of clusters versus cluster size n are shown in Fig. 9. The number of secondary carbon atoms in the pure $C$ clusters in $C_{18}$, $C_{25}$, $C_{33}$, and $C_{43}$ are more than their neighbours. This leads to an increase of the number of atoms with higher energies.

Thus, the abundance of secondary carbon atoms is the cause of the structure weakness.

Generally, $H$ passivated clusters have more secondary carbon atoms resulting in an increase of the number of ternary carbon atoms. Those atoms have a lower binding energy as compared to the case of pure $C$ clusters. In all cases when the clusters have a smaller number of secondary carbon atoms, they have a larger binding energy (see Fig. 2).

\section{TRIGONAL AND HEXAGONAL SHAPED NANODISKS}

Highly symmetric nanodisks have been recently the focus of theoretical research\cite{ezaw,fern,pota}. Such nanodisks are nanometer scale disk-like graphene materials which have closed edges. In particular the electronic structure of nanodisks that have the same type of edges, i.e. zigzag or armchair, have been investigated. Here we limit ourselves to trigonal shaped nanodisks that have zigzag edges\cite{you} and hexagonal shaped structures\cite{zhang}	that can have either zigzag or armchair edges (see Fig. 10). The interest in those nanodisks originates from the prediction that zigzag terminated nanodisks have degenerate zero-energy electronic states\cite{ezaw,zhang} that leads to metallic ferromagnetism\cite{fern,pota}. Recently, STM measurements have given evidence for the existence of metallic zero-gap graphene flakes with zigzag termination\cite{ritt}.

\begin{figure} \vspace*{0.9cm}\centering {\resizebox*{!}{6.0cm}
{\includegraphics{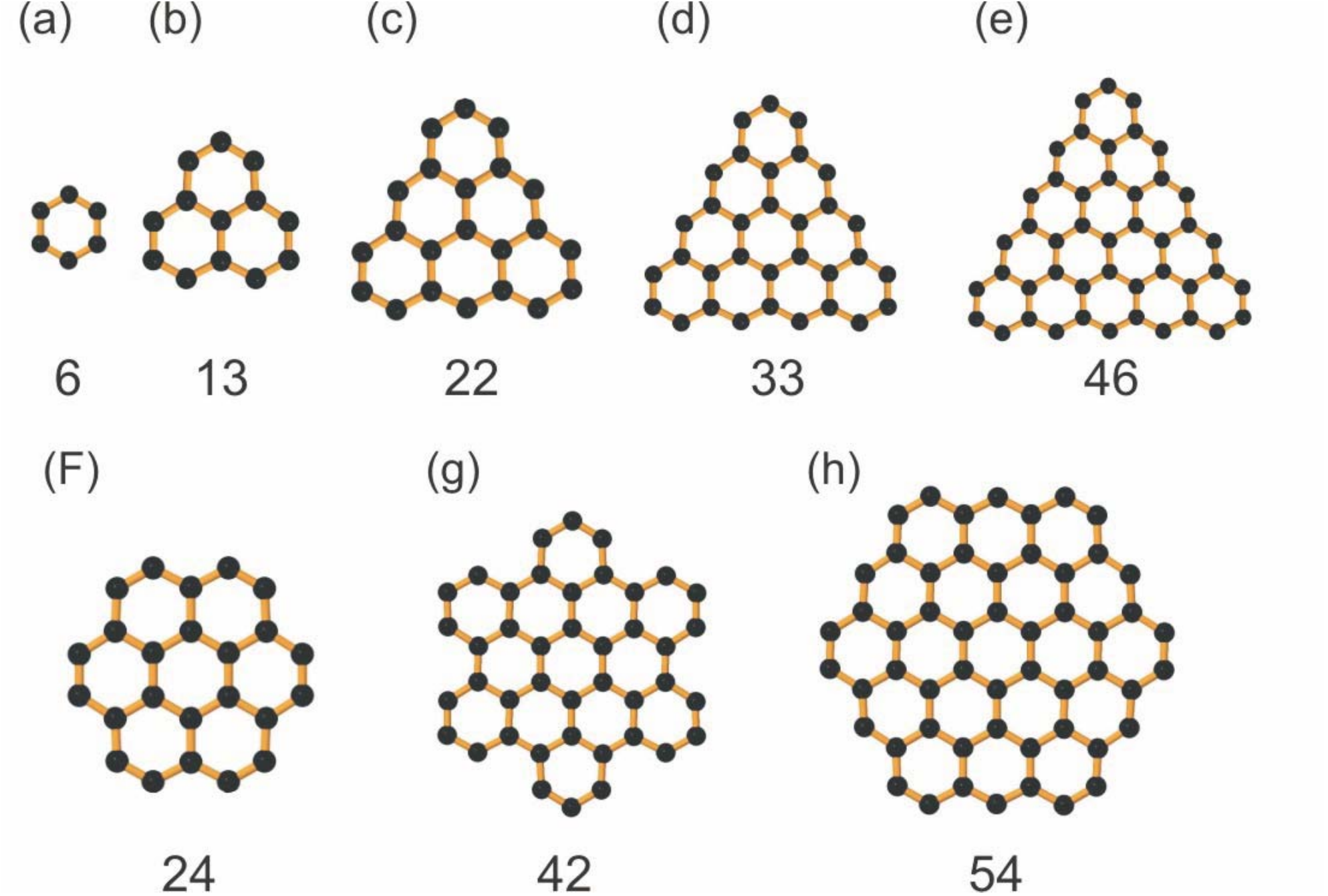}}} \caption{
Nanodisks with zigzag edges for trigonal and with zigzag and/or armchair edges for hexagonal shaped structures.}\label{fig12}
\end{figure}

We show here that these interesting and promising graphene nano-islands are not necessarily the ground state configuration for the given number of C-atoms and in such a case will be metastable. Nevertheless, it is possible to realize such stable and metastable graphene nanostructures with zigzag or armchair edges through Joule heating of nanoribbons\cite{ritt2}, electron bombardment\cite{yiri} or by crystallographic specific etching techniques\cite{datt} of a graphene layer.

In the theoretical investigation of nanodisks one assumes that they are build up with hexagonal plaquettes. This does not necessarily lead to the most energetic favorable state and sometimes the insertion of pentagons and/or heptagons may lower their energy. A typical example is the buckyball. We showed already that it is not necessary to have a three-dimensional structure for the occurrence of non-hexagonal polygons and that even a perfect planar structure can have such polygons.

We consider the trigonal and the two hexagonal shaped clusters depicted in Fig. 10. The trigonal structure has zigzag edges with a total number of $n=1+4N_{side}+N^2_{side}$ carbon atoms where $N_{side}$ is the number of hexagonal plaquettes along each side of the triangle. For $N_{side}$=1, 2, ..., 5 we have $n$=6, 13, 22, 33, 46 respectively. The first one, i.e. $n$=6, is a single hexagon which was found to be the ground state (see Table I). In Table II we also show the energy of the unrelaxed cluster where the C-C distances are taken equal to the bulk value, a=1.42 \r{A}. Notice that the relaxation of the configuration, in particular of the edge atoms, lowers the energy with about 0.4\%. All trigonal shaped structures shown in Fig. 10 with the exception of $n$=6 are metastable state configurations (see Table II).	

The hexagonal nano-island with zigzag edges has	$n=6N^2_{size}$ carbon atoms. For $N_{side}$=1, 2, 3 we have respectively $n$=6, 24, 54. We found that all these 3 configurations are ground state structures. This type of clusters are perfectly close-packed planar structures with purely zigzag edges.

\begin{table}[htdp]
\caption{The binding energy per atom (in eV/atom) as function of the number of sides in the cluster $(N)$ and $n$ is the total number of C-atoms. We compare the ground state $(GS)$ energy per atom of the pure carbon clusters with the H-passivated clusters. For the pure carbon clusters we also present the results for the unrelaxed situation ($nr$).}
\begin{center}
\resizebox{\linewidth}{!}{%
\begin{tabular*}{\linewidth}{lcccccc}
\hline\hline
 $N_{size}$ &  ~n & ~~~$E_b^{nr}$ & ~~~$E_b$ & ~~~$E_b(GS)$ & ~~~$E_b$ & ~~~$E_b(GS)$ \\ 
& & ~~~(C-C) & ~~~(C-C) & ~~~(C-C) & ~~~(C-H) & ~~~(C-H) \\
\hline
\multicolumn{7}{l}{Trigonal}\\
1 &    6 & ~-5.2682 & ~-5.2869 & ~-5.2869 & ~-7.6233 & ~-7.6233 \\
2 & 13 & ~-5.8115 & ~-5.8427 & ~-5.9349 & ~-7.5557 & ~-7.5557 \\
3 & 22 & ~-6.1036 & ~-6.1339 & ~-6.1674 & ~-7.5219 & ~-7.5429 \\
4 & 33 & ~-6.2969 & ~-6.3251 & ~-6.3373 & ~-7.5009 & ~-7.5010 \\
5 & 46 & ~-6.4371 & ~-6.4628 & ~-6.5195 & ~-7.4862 & ~-7.4864 \\
\multicolumn{7}{l}{Hexagonal (zigzag edges)}\\
1 &   6 & -5.2682 & -5.2869 & -5.2869 & -7.6233 & -7.6233 \\
2 & 24 & -6.0382 & -6.2368 & -6.2368 & -7.5121 & -7.5121 \\
3 & 54 & -6.4534 & -6.5973 & -6.5973 & -7.4730 & -7.4815 \\
\multicolumn{7}{l}{Hexagonal (armchair edges)}\\
1 &   6 & -5.2682 & -5.2869 & -5.2869 & -7.6233 & -7.6233 \\
2 & 42 & -6.2737 & -6.4628 & -6.4913 & -7.4943 & -7.4943 \\[1ex]
\hline
\end{tabular*}}
\end{center}
\label{table:nonlin}
\end{table}%

The hexagonal nano-island with armchair edges has $n=6(1- 3N_{side}+3N^2_{side})$ carbon atoms. The $C_{42}$ structure is a metastable state cluster, while the ground state is a close-packed graphene- type structure (see Fig. 1).

As we have noted above, the trigonal structures have zigzag edges, and they are metastable configurations when they consist of only carbon atoms. We studied those clusters with $H$ atoms attached to their edges, and found that $C_6$, and the $C_{13}$ structure are ground state, and $C_{22}$, $C_{33}$ and $C_{46}$ $H$ passivated structures are metastable clusters.

The trigonal shaped structures have $N_s=(6+4N_{side})-N_{side}$ secondary carbon atoms. The $(C-C)_{22}$, $(C-H)_{22}$, $(C-C)_{33}$ and $(C- H)_{33}$ clusters have the same number of secondary carbon atoms $N_s$=12 and $N_s$=15 respectively. $(C-C)_{13}$ is a single ring structure, while $(C-H)_{13}$ is a graphene-type cluster. In the $(C-H)_{13}$ structure all $N_s$=9 secondary carbon atoms after $H$ passivation of the edge atoms become ternary atoms.
W e	observed,	that	the	small	$(C-H)_{24}$	hexagonal	shaped clusters with zigzag edges is the ground state structure, while $(C-H)_{54}$ becomes a metastable configuration. $(C-H)_{42}$ is the ground state structure because of the abundance of secondary carbon atoms.

\section{CONCLUSIONS}

Atomic simulations where used with the Brenner second generation reactive bond order (REBO) inter-atomic potential function to study small neutral carbon clusters $C_n	(11\le n \le 55)$. In the present work we are interested in quasi-planar structures and therefore omitted the conjugate compensation term $F_{ij}(N_i, N_j, N^{conj}_{ij})$ of the original potential function. We concentrated on the ground state configurations. Our results show that $C_n$ is a single ring for $n$ up to 18, and multi-rings for $19\le n \le 55$. We found that $C_{20}$, $C_{28}$, $C_{38}$, and $C_{44}$ are buckled clusters with a pentagon in the center.

The binding energy of the clusters does not exhibit even-odd number oscillations. But we found smaller energy oscillations which are due to the occurrence of special type of structures, and in particular the kind of edge they have (i.e. zigzag or armchair type of edges). It also depends on the position of some of the polygons in the cluster. If the polygons are far removed from each other, then the cluster has a higher energy, and if the polygons are closer to each other the cluster has a lower binding energy. Similarly, we found that clusters with mostly zigzag edges have lower binding energy, especially if they are �close- packed� structures than those consisting of both: zigzag and armchair edges.

The hexagons play an essential role in the formation of graphene-like clusters. Pentagons can be in the center of the cluster, as well as at the edge, while heptagons were never found at the center of the ground state structure. Clusters having the graphene structure are not bended or rippled but are perfectly flat.

Our results (see Fig. 2(a)) for e. g. the energy, are different from previous works which we can trace back to the use of a different inter-atomic potential function. Wang et al.\cite{wang} and Zhang et al.\cite{zhang} used the previous version of the Brenner inter- atomic potential function which differs from ours in the fact that our (REBO) potential function contains improved analytic functions and was based on an extended database as compared to the early version (Refs. 28, 42). The bond energies, lengths, and force constants for hydrocarbon molecules are significantly better described in the second version of the potential.

We found that trigonal shaped clusters having zigzag edges are metastable structures. The hexagonal shaped clusters that have only zigzag type edges are close-packed ground state structures and are always the ground state (except for $H$ passivated clusters), while the same shaped clusters that have only armchair type edges are metastable structures (see Table II). These conclusions are valid except for the trivial case $n=6$.

Our calculations show that some of the metastable carbon clusters can become the ground state configuration if H-atoms are attached to the edge atoms. The number of pentagons influence the binding energy of those $H$ passivated clusters. Clusters with pentagons have higher energy. The binding energy of $H$ passivated clusters is larger in comparison to the pure carbon clusters but both converge to the graphene result in the limit $n\to \infty$.

For $n>11$ graphene-type of clusters can be formed as ground state. This is even more so for $n>41$.

\begin{acknowledgements}
 This work was supported by the Belgian Science Policy (IAP) and the Flemish Science Foundation (FWO-Vl).
\end{acknowledgements}

\end{document}